# Quantum revivals and magnetization tunneling in effective spin systems

**M. Krizanac[1], D. Altwein[1], E. Y. Vedmedenko[1] and R. Wiesendanger[1]**

[1] University of Hamburg, Institute for Applied Physics, Jungiusstr. 11a, 20355 Hamburg

E-mail: mkrizana@physnet.uni-hamburg.de

**Abstract.** Quantum mechanical objects or nanoobjects have been proposed as bits for information storage. While time-averaged properties of magnetic, quantum-mechanical particles have been extensively studied experimentally and theoretically, experimental investigations of the real time evolution of magnetization in the quantum regime were not possible until recent developments in pump-probe techniques. Here we investigate the quantum dynamics of effective spin systems by means of analytical and numerical treatments. Particular attention is paid to the quantum revival time and its relation to the magnetization tunneling. The quantum revival time has been initially defined as the recurrence time of a total wave-function. Here we show that the quantum revivals of wave-functions and expectation values in spin systems may be quite different which gives rise to a more sophisticated definition of the quantum revival within the realm of experimental research. Particularly, the revival times for integer spins coincide which is not the case for half-integer spins. Furthermore, the quantum revival is found to be shortest for integer ratios between the on-site anisotropy and an external magnetic field paving the way to novel methods of anisotropy measurements. We show that the quantum tunneling of magnetization at avoided level crossing is coherent to the quantum revival time of expectation values, leading to a connection between these two fundamental properties of quantum mechanical spins.

## 1. Introduction

The wave-function associated with a particle never disappears completely. Therefore it is not surprising that under certain conditions, for instance application of a periodic force, the wave-function of such a particle returns to its initial state. The wave-function of a quantum particle can decay over time, but its initial state is not lost and can reappear in certain time windows. This phenomen of the wave-function reincarnation is known as the quantum revival [1, 2, 3, 4]. Recovering the complete wave-function destroyed by the decay processes has fascinated researches in the context of two-level quantum systems [5, 6, 7], quantum wells [8, 9, 10, 11, 12] and Bose-Einstein condensates [13, 14]. Meanwhile, the quantum dynamical phenomena become increasingly important also in solid state physics. Particularly, quantum mechanical tunneling or the non-classical field dependence of magnetization has been reported for nanomagnets [15], molecules [16, 17] and single atoms [18]. Therefore, instead of studies of time-averaged properties like, e.g., magnetization curves [18, 19], nowadays the emphasis is put on the time dependent behavior of magnetization [20, 21, 22, 23, 24]. This trend is promoted further by the development of novel pump-probe techniques [26] allowing for the sub-femto-second time resolution of magnetization dynamics, which might shed light on the revival phenomena in nano-magnetic systems. Hence, theoretical predictions are urgently required. Until now, however, the systematic theoretical description of quantum revival in magnetic systems is lacking, while existing investigations come to controversial conclusions. Particularly, an increase as well as decrease of the quantum revival time (QRT) with increasing spin $s$ has been reported [20, 21].
Here, an exact expression for the time evolution of an effective quantum magnetic moment has been derived analytically using the Schrödinger formalism. In contrast to previous investigations, which do not differentiate between the revival of the total wave function (state) and that of the expectation values of the wave-function, we distinguish the quantum revival time of the total wave-function (QRT) and the revival time of expectation values (EVRT) and show that they are not identical. We concentrate on EVRT as only this quantity can be measured experimentally. The analysis of the expression, which we obtain for EVRT, shows that the time-dependent behavior of spin operators can be represented via the Fourier series of characteristic frequencies $\omega_i$. These frequencies, in turn, define the non-harmonic precession of expectation values of spin components. Surprisingly, it doesn't depend on the angular momentum as previously predicted, but rather is defined via the ratio of the anisotropy constant $K$ and the external magnetic field $B$. The shortest EVRT can be found for $\widetilde{B}_z = N \cdot K$ with any integer $N$. For any other $\widetilde{B}_z/K$ ratio the EVRT is larger or even infinite for an irrational ratio. Furthermore, our analysis reveals that the quantum tunneling of magnetization occurs at $\widetilde{B}_z = N \cdot K$, where $N \leq 2s$, in the regime of small transversal fields. Hence, the QRT and the magnetization tunneling are closely related to one another.

## 2. Time evolution of expectation values

The Quantum Recurrence Theorem [25] defines the QRT as the shortest time interval $\Delta t$ after which the full wave-function $\big|\Psi(t)\big\rangle$ of a system periodically repeats $\big|\Psi(t)\big\rangle = \big|\Psi(t+\alpha\Delta t)\big\rangle$ $\alpha \in \mathbb{N}$. In experiment, however, only measurements of expectation values of the magnetization components $\big\langle S_j\big\rangle_t$ ($j := x, y, z$) are possible. Consequently, the periodicity of $\big|\Psi(t)\big\rangle$ recurrence is inaccessible, while that of the expectation values can be measured. The periodicity of expectation values may be different from the standard QRT as has been shown for a particle in a quantum well [12]. However, systematic comparison of these two quantities in quantum magnetic systems is still lacking.
For the sake of a systematic analysis of the dynamics of expectation values the time-dependent equation of motion

$$i\hbar\frac{\partial}{\partial t}\Psi(t) = \hat{H}\Psi(t) \tag{1}$$

has been solved analytically. The Hamilton operator in (1) includes the standard spin operators $\hat{S}_i$, uniaxial on-site anisotropy $K$ and a magnetic field $\vec{B} = (0, 0, B_z)$. This Hamiltonian models an effective quantum spin representing a nanoparticle, a molecule or an atomic cluster [27, 19, 15].

$$\hat{H} = -\hat{S}_z\widetilde{B}_z - K\hat{S}_z^2 \quad , \tag{2}$$

where

$$\widetilde{B}_z = \mu_B B_z. \tag{3}$$

The equation (1) and (2) define a system of uncoupled linear ordinary differential equations of first order

$$i\hbar\frac{\partial}{\partial t}\Psi(t) = \begin{pmatrix} -s\widetilde{B}_z - s^2K & \cdots & 0 \\ \vdots & \ddots & \\ 0 & & s\widetilde{B}_z - s^2K \end{pmatrix} \cdot \begin{pmatrix} \varphi_{+s}(t) \\ \vdots \\ \varphi_{-s}(t) \end{pmatrix} \tag{4}$$

with the following solutions for the eigenstates

$$\big|\Psi(t)\big\rangle = \begin{pmatrix} \varphi_{+s}(t) = \varphi_{+s}(t_0)\cdot e^{-i(-s\widetilde{B}_z - s^2K)t/\hbar} \\ \vdots \\ \varphi_{-s}(t) = \varphi_{-s}(t_0)\cdot e^{-i(s\widetilde{B}_z - s^2K)t/\hbar} \end{pmatrix} \tag{5}$$

where $s$ is the spin quantum number and the initial conditions are

$$\varphi_{+s}(t_0), ..., \varphi_{-s}(t_0) \in \mathbb{C}. \tag{6}$$

The normalization condition is given by the expression

$$\sum_{i=0}^{2s} |\varphi_{s-i}(t)|^2 = |\varphi_{+s}(t)|^2 + \ldots + |\varphi_{-s}(t)|^2 = 1. \tag{7}$$

With the knowledge of eigenstates, the time-dependent expectation values of spin operators can be obtained as

$$\langle \hat{S}_i \rangle_t = \langle \Psi(t) | \hat{S}_i | \Psi(t) \rangle \qquad i := x, y, z. \tag{8}$$

An example of a real part of the expectation value for the spin operator $\hat{S}_x$ with $s = 1\hbar$ is given by (9).

$$\left\langle S_{x_{real}} \right\rangle_t = \frac{2\hbar}{\sqrt{2}} \cdot \Big[ (\underset{real}{\varphi_{+1}(t_0)} \underset{real}{\varphi_0(t_0)}) \cdot \cos\big((-\widetilde{B}_z - K)\frac{t}{\hbar}\big) + (\underset{real}{\varphi_{-1}(t_0)} \underset{real}{\varphi_0(t_0)}) \cdot \cos\big((\widetilde{B}_z - K)\frac{t}{\hbar}\big) \Big]. \tag{9}$$

The complete expression for the expectation values of $s = 1\hbar$ can be found in Appendix A. A remarkable peculiarity of the expression (9) is the superposition of two characteristic frequencies $\omega_1 = (-\widetilde{B}_z - K)/\hbar$ and $\omega_2 = (\widetilde{B}_z - K)/\hbar$, which appear due to the quadratic nature of the uniaxial anisotropy term in the Hamilton operator of (2). The presence of two harmonics in (9) defines a non-harmonic oscillation [30], which is characterized by time-dependent amplitudes.

The generalization of expressions (B.1), (9) for any spin value is derived in Appendix B. It reveals that for a Hamilton operator including anisotropy terms with an even exponent, for example, $K \cdot \hat{S}_\alpha^{2n}$ ($\alpha = x, y, z$ and $n \in \mathbb{N} \setminus \{0\}$), the expectation values $\left\langle S_j \right\rangle_t$ ($j := x, y$) can be represented in the form of Fourier series, while $\left\langle S_z \right\rangle_t$ is a constant. A solution of $\left\langle S_x \right\rangle_t$ for an arbitrary value of a quantum spin number is given by (10), while the solution for $\left\langle S_y \right\rangle_t$ can be found in Appendix B:

$$\left\langle S_x \right\rangle_t = \sum_{i=1}^{2s} \big( \alpha_i \cdot \cos(\omega_i t) + \beta_i \cdot \sin(\omega_i t) \big) \tag{10}$$

where

$$\begin{aligned} \alpha_i &= 2(S_x)_{i,i+1} \Big( \varphi(t_0)^{real}_{s-(i-1)} \cdot \varphi(t_0)^{real}_{s-i} + \varphi(t_0)^{imag}_{s-(i-1)} \cdot \varphi(t_0)^{imag}_{s-i} \Big) \\ \beta_i &= 2(S_x)_{i,i+1} \Big( -\varphi(t_0)^{real}_{s-(i-1)} \cdot \varphi(t_0)^{imag}_{s-i} + \varphi(t_0)^{imag}_{s-(i-1)} \cdot \varphi(t_0)^{real}_{s-i} \Big) \end{aligned} \tag{11}$$

and

$$(S_x)_{i,i+1} = \frac{\hbar}{2} \sqrt{(s+1)2i - (i^2 + i)} \tag{12}$$

$$\omega_i = \frac{\big( -\widetilde{B}_z - (2s - (2i - 1))K \big)}{\hbar}. \tag{13}$$

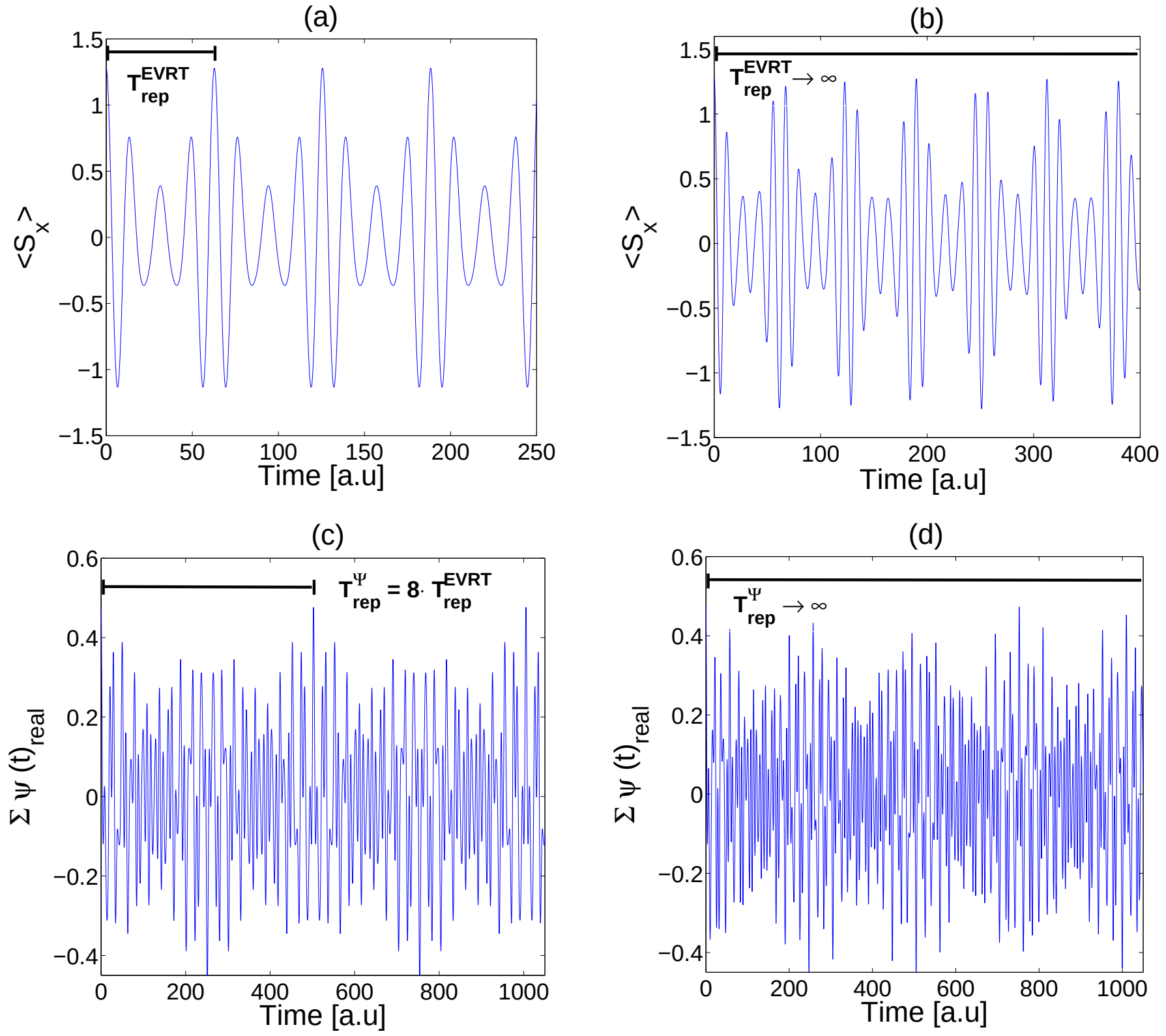

**Figure 1.** Example of the time-evolution for an effective quantum spin $3\hbar/2$: **a)** The time evolution of the expectation value $\langle S_x \rangle_t$ for $\widetilde{B}_z = 0.5$ [a.u] and $K = 0.05$ [a.u] ($\widetilde{B}_z/K = 10$). The lowest possible periodicity of the expectation value corresponds to $T_{rep} \approx 62$ [a.u]; **b)** The same $\langle S_x \rangle_t$ for $\widetilde{B}_z = 0.18 \cdot \pi$ [a.u] and $K = 0.05$ [a.u], with $\widetilde{B}_z/K \in \mathbb{I} = \mathbb{R} \backslash \mathbb{Q}$ leads to $T_{rep} \rightarrow \infty$; **c)** Time evolution of the total wave-function $|\Psi(t)\rangle$. The $\Sigma\psi(t)_{real}$ represents the real part of the total wave-function components. For the same magnitude of $\widetilde{B}_z$ and $K$ as in (a), much larger values of $T_{rep} \approx 500$ [a.u] compared to (a) have been obtained; **d)** $\Sigma\psi(t)_{real}$ for parameters used in (b) with $T_{rep} \rightarrow \infty$.

Hence, the time evolution of the magnetization components of an arbitrary quantum spin subject to quadratic terms in the Hamilton operator can be represented by a superposition of $2s$ characteristic frequencies $\omega_i$. Clearly, these frequencies define a general non-harmonic oscillation. An exception is given by a spin $\hbar/2$ particle as in this case only one frequency $\widetilde{B}_z/\hbar$ survives. The spectrum of a general non-harmonic oscillation is discrete and is characterized by its fundamental frequency $\omega_f$. The fundamental frequency $\omega_f$ corresponds to the greatest common divisor (gcd) of all harmonics $\gcd(\omega_1, ..., \omega_i)$. In our case all these frequencies depend on the magnitudes of $K$ and $\widetilde{B}_z$. In the simplest case of spin $1\hbar$ there are only two of them, namely $\omega_1 = (-\widetilde{B}_z - K)/\hbar$ and $\omega_2 = (\widetilde{B}_z - K)/\hbar$. Quite generally, the shortest period of any oscillation corresponds to a reciprocal of the fundamental frequency defined above and is known as the repetitive time $T_{rep} = \frac{2\pi}{\omega_f}$

$$T_{rep} = \frac{2\pi}{\gcd(\omega_1, ..., \omega_i)} \quad , \ \omega_i \in \mathbb{Z} \tag{14}$$

In figure 1(a, c) a portion of the time dependent evolution of $\langle S_x \rangle$ and the total wave-function with integer $\widetilde{B}_z/K$ value are shown, while in Fig. 1 (b, d) the signal for the case of an irrational $\widetilde{B}_z/K$ ratio is presented. While both cases represent the non-harmonic oscillation their behavior shows significant differences. Most importantly, $\langle S_x \rangle_t$ for $\widetilde{B}_z/K = 10$ of Fig. 1 (a) shows much smaller periodicity $T_{rep}$, than that of irrational $\widetilde{B}_z/K$ (Fig. 1 (b)). In figures 1(c) and 1(d) which represent the time evolution of the sum $\Sigma\psi(t)$ of the wave-function $|\Psi(t)\rangle = (\psi_s(t), ..., \psi_{-s}(t))^T$ the same values of $\widetilde{B}_z$ and $K$ have been used. Nevertheless, much longer repetition periods of expectation values have been obtained. For irrational $\widetilde{B}_z/K$ ratio the periodicity of the total wave-function is infinite. In the next step, the relation between the repetitive time, the QRT and the EVRT for a given Hamilton operator will be revealed.

## 3. Non-harmonic revival of expectation values

As the repetitive time gives the shortest period of oscillation it corresponds to the revival time. Figure 1 demonstrates that the $T_{rep}$ for the time evolution of wave-functions (Figure 1(c,d)) and that of expectation values (Figure 1(a,b)) may be quite different even for the same Hamiltonian. Additionally, the frequencies $\omega_i$ which determine the $T_{rep}$ depend on the ratio of the external magnetic field and the anisotropy. In the next step the reasons for differences between QRT and EVRT as well as the role of the $\widetilde{B}_z/K$ ratio for $T_{rep}$ will be analyzed.

While expectation values are always real, the wave-functions and coefficients $\alpha_i$ in (10) may be complex. For the sake of comparison between EVRT and QRT analytically we first explore the real solution for $T_{rep}$ only. However, the generalization to complex variables is straightforward and doesn't change our conclusions. For that purpose we use the Bézout's identity [29] stating:

$$\gcd(\omega_1, ..., \omega_m) = \sum_{i=1}^{m} a_i \cdot \omega_i \tag{15}$$

where $a_i, \omega_i \in \mathbb{Z}$. With (15) $T_{rep}$ becomes

$$T_{rep} = \frac{2\pi}{\gcd(\omega_1, ..., \omega_m)} = \frac{2\pi}{a_1\omega_1 + ... + a_m\omega_m}. \tag{16}$$

The numerator and denominator of this expression can be expanded as

$$T_{rep} = \frac{2\pi \cdot 10^n}{a_1\omega_1 10^n + ... + a_m\omega_m 10^n} = \frac{2\pi \cdot 10^n}{\gcd(\omega_1 \cdot 10^n, ..., \omega_m \cdot 10^n)}, \tag{17}$$

where $\omega_i \in \mathbb{R}$, $n \in \mathbb{N}$ and $\omega_i \cdot 10^n \in \mathbb{Z}$. The expansion of (17) leaves the period $T_{rep}$ unchanged and Bézout's identity unaffected. It means that the initial greatest common divider can be replaced by an equivalent expression using simple multiplication of the characteristic frequencies and the numerator by $10^n$. For any irrational number $\widetilde{B}_z/K \in \mathbb{R}\backslash\mathbb{Q}$ the sequence in the denominator of (17) and, hence, $T_{rep}$ is infinite. This fact explains why we were not able to find the quantum revival in figure 1(b,d). For any integer or rational $\widetilde{B}_z/K \in \mathbb{Q}$, the $T_{rep}$ is finite. Hence, the interesting question arises how the revival time of expectation values is connected to the $\widetilde{B}_z/K$ ratio in this case. This question is addressed in detail in Appendix C.

The fundamental frequency $\omega_f = \gcd(\omega_1 \cdot 10^n, ..., \omega_m \cdot 10^n)$ of EVRT is always equal or lower than the lowest $\omega_i$ in (10), which can be expressed by an infimum (inf) :

$$\omega_f \leq \inf(\omega_1 \cdot 10^n, ..., \omega_m \cdot 10^n). \tag{18}$$

For a given spin value $s$ the allowed frequencies of individual harmonics $\omega_i$ given in (13) depend on the magnitude of $\widetilde{B}_z$ and $K$ only. Defining $\widetilde{B}_z/K = N$ for a given $s$ and replacing $\widetilde{B}_z$ by $N \cdot K$ in (13) we end up with the following expression:

$$\omega_i = \frac{\big[-N-(2s-(2i-1))\big] \cdot K}{\hbar} \tag{19}$$

which leads to:

$$\omega_i = \begin{cases} (-N-1)K, (-N-3)K, ..., (-N-(2n+1))K & \text{for integer } s \\ (-N)K, (-N-2)K, (-N-4)K, ..., (-N-2n)K & \text{for half-integer } s \end{cases} \tag{20}$$

with $n$-number of harmonics. Hence, the spin value $s$ is not at all present in the analytical expression (20). As $i$ is an integer and $s$ is an integer or half-integer in (19) the fundamental frequency corresponds to the very specific choices of $\widetilde{B}_z/K = N$. All possible values of the fundamental frequency $\omega_f$ for integer, rational or irrational N and different spin numbers are derived in Appendix C. The most important results are summarized in the following.

Most importantly, the $T_{rep}$ corresponding to the fundamental frequency $\omega_f$ is different for different combinations of spin statistics (integer or half-integer) and $N$ ratios. The lowest possible $T_{rep}$ and, hence, the lowest EVRT among all possible combinations of $N$ and the spin statistics appears for any

$$\frac{\widetilde{B}_z}{K} = N \in \mathbb{Z}. \tag{21}$$

Particularly,

$$T_{rep} = \begin{cases} \dfrac{2\pi\hbar}{K}, & \text{(integer } s \text{ and even } N \text{ or } N=0) \text{ or (half-integer } s \text{ and odd } N) \\ \dfrac{\pi\hbar}{K}, & \text{(integer } s \text{ and odd } N) \text{ or (half-integer } s \text{ and even } N \text{ or } N=0) \end{cases} \tag{22}$$

Thus, there are only two possible values of the revival time for the expectation values $T_{rep}^{\mathrm{EVRT}} = \frac{2\pi\hbar}{K} \vee \frac{\pi\hbar}{K}$ for any $N \in \mathbb{Z}$. These EVRT do not depend on the spin values as predicted before[20, 21], but only on the spin statistics. Because of our analytical findings it becomes clear why it happens. As the spin dynamics is described by the superposition of harmonic frequencies there is only one fundamental frequency for the given set of parameters.

If $N$ is rational $\widetilde{B}_z/K = N \in \mathbb{Q}/\mathbb{Z}$, for example $N = 3/5$, the revival times are still finite but always somewhat larger than those in case of $N \in \mathbb{Z}$,

$$T_{rep} = \begin{cases} \dfrac{\pi b}{K} & , b > 2 \\ \dfrac{2\pi b}{K} & , b > 1 \end{cases} \tag{23}$$

where $b$ is the denominator from the definition of $N = a/b \in \mathbb{Q}$ in Appendix C. In summary,

$$T^{\text{EVRT}}_{N\in\mathbb{R}\backslash\mathbb{Q}} > T^{\text{EVRT}}_{N\in\mathbb{Q}\backslash\mathbb{Z}} > T^{\text{EVRT}}_{N\in\mathbb{Z}} \tag{24}$$

where $\mathbb{R}\backslash\mathbb{Q}$ are irrational numbers. The derivation of (C.21) and (24) can be found in Appendix C. An external field can easily be tuned experimentally. Therefore one can use the time dependent measurement of EVRT to determine the anisotropy of the system. It can be done by experimental measurements of the revival times. Knowing the value of the field at the shortest revival the anisotropy value can be derived using (C.6).

Up to this point only the EVRT has been considered. The QRT have been addressed in detail in Appendix D using the procedure similar to that of Appendix C. We find that EVRT and QRT are identical for all effective quantum spins with integer spin number, while they are different for half-integer $s$. Particularly,

$$\begin{aligned} T^{\text{EVRT}}_{rep} &= T^{\text{QRT}}_{rep}, && \text{for integer } s \\ T^{\text{EVRT}}_{rep} &= \frac{1}{\alpha}\cdot T^{\text{QRT}}_{rep}, \quad \alpha \in [1,2,4,8], && \text{for half-integer } s \end{aligned} \tag{25}$$

In contrast to EVRT for half-integer spins, $T^{\text{QRT}}_{rep}$ is shortest for $\widetilde{B}_z/K = (2\beta-1)/2 \in \mathbb{Q}\backslash\mathbb{Z}$ with $\beta \in \mathbb{Z}$. This means, that while the expectation values of magnetization repeat fastest for any $\widetilde{B}_z/K \in \mathbb{Z}$ it is not the case for the wave-functions as $(2\beta-1)/2 \notin \mathbb{Z}$.

In summary, the standard QRT and the EVRT of an effective spin are identical for even spin numbers but very different for half-integer spins. The EVRT depends on the ratio $N$ only but is independent of the magnitude of the spin quantum number. The shortest revivals can be observed for $N \in \mathbb{Z}$.

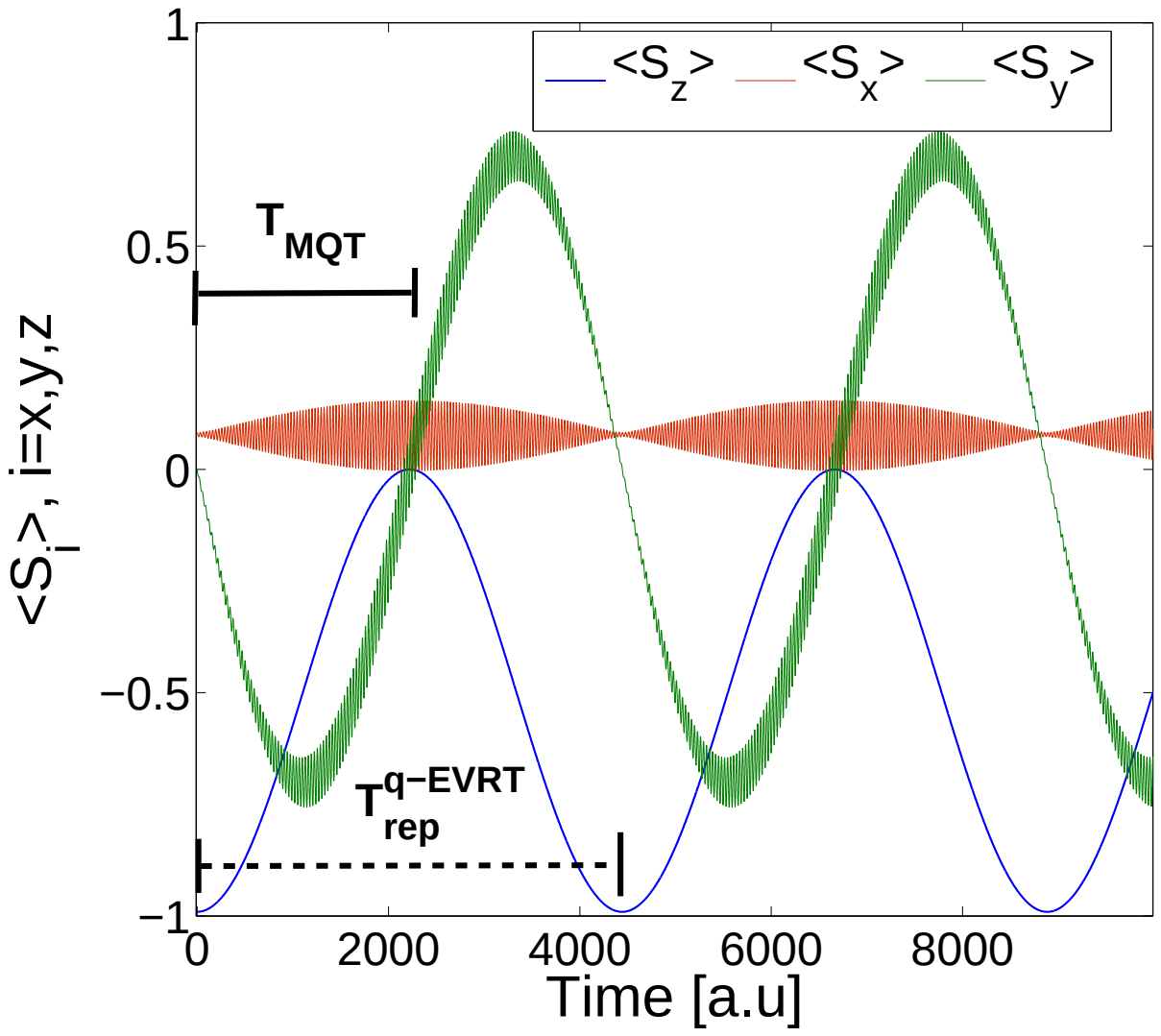

**Figure 2.** Behavior of expectation values within the tunneling regime obtained using the numerical solution of the Schrödinger equation. For $\widetilde{B}_z/K = N \in \mathbb{Z}$ with $\widetilde{B}_z = 0.1$ [a.u], $K = 0.1$ [a.u] and a small transversal field $\widetilde{B}_x = 0.001$ [a.u]. The repetitive time $T_{rep}$ is equal for the expectation values $\langle S_j \rangle_t$ $(j := x, y, z)$ and it is two times the magnetization tunneling time $T_{rep}^{\text{q-EVRT}} \approx 2 \cdot T_{MQT}$.

## 4. Quasi-Quantum revival and magnetization tunneling

Previous considerations of quantum spin dynamics in the literature have been performed for uncoupled and unperturbed systems [20, 21]. In the following we want to analyze the behavior of the EVRT under the influence of a small time-independent perturbation. Such an investigation is important, because small perturbations like a transversal field or anisotropy break the rotational symmetry about the z-axis and make the $\hat{S}_z$ spin operator and the energy non-commuting. This symmetry breaking promotes the spontaneous transition of magnetization from one energy well (expectation value of magnetization) to another one under the energy barrier, known as the resonant quantum tunneling of magnetization (MQT). The MQT have been measured in many solid state and molecular magnetic systems and is of large technological importance [17, 24, 18, 27, 22], because it is directly connected to the life-times of the information bits. In zero external magnetic field $\vec{B}_z = 0$, the quantum mechanical levels $m = \pm s$ have the lowest energy. When a non-zero $\vec{B} = (0, 0, B_z)$ is applied, the levels with $m > 0$ decrease in energy, while those with $m < 0$ increase. At certain $B$ values the crossing of positive and negative levels occurs. When a transversal field is added the avoided level crossing appears[17, 27, 31]. This avoided level crossing promotes the MQT. However, the probability of the MQT depends strongly on the velocity at which the field $B(t)$ is sweeped [31, 32]. Hence, it is reasonable to anticipate that some connection between the EVRT and the probability of MQT exists. In the following this connection will be explored.

The interplay between EVRT and QRT has been studied by the numerical solution of (1). We used Simpson's rule, which is similar to the Runge-Kutta procedure of third order for the numerical approximation of the solution of (1). The Hamiltonian (2) has been modified to

$$\hat{H} = -\hat{S}_z\widetilde{B}_z - K\hat{S}_z^2 - \hat{S}_x\lambda\widetilde{B}_x \tag{26}$$

with $\lambda\widetilde{B}_x \ll \widetilde{B}_z$. Figure C1 gives time evolution of $\langle S_z \rangle$, $\langle S_x \rangle$ and $\langle S_y \rangle$ values of $s = 1\hbar$ for $\widetilde{B}_z/K = 1$. In contrast to figure 1, where $\langle S_z \rangle$ was always constant, the vertical component of magnetization in the present case oscillates between $\langle S_z \rangle = -1$ and $\langle S_z \rangle = 0$ values, while the $y$-component of magnetization switches (tunnels) between $\langle S_y \rangle \approx -0.7$ and $\langle S_y \rangle \approx 0.7$. This behavior is typical for the MQT. The time, which is needed to switch the magnetization $\langle S_z \rangle$ from the $|-1\rangle$ to the $|0\rangle$ state is denoted in the following as magnetization tunneling time $T_{MQT}$.

The numerical analysis of the results obtained for several $N$-ratios has shown that for $\widetilde{B}_x \ll \widetilde{B}_z$ the $T_{rep}^{\mathrm{EVRT}}$ is of the order of $2 \cdot T_{MQT}$. This definition is not mathematically exact, because, strictly speaking, the EVRT cannot be described analytically for perturbed arbitrary quantum spin systems. However, the numerical data indicates that the MQT oscillation (see $\langle S_z \rangle_t$ in figure C1), with small $\widetilde{B}_x$, is similar to an harmonic oscillation. This means we describe for small values of $\widetilde{B}_x$ a non-harmonic oscillation as a harmonic oscillation, because the differences $\Delta_{<S_z>} = \langle S_z \rangle_{\mathrm{minima}_1} - \langle S_z \rangle_{\mathrm{minima}_2}$ between two successive minima of $\langle S_z \rangle_t$ decrease for decreasing $\widetilde{B}_x$ and the non-harmonic behavior is not immediately obvious.

$$\lim_{\widetilde{B}_x \to 0} \Delta_{<S_z>} = \lim_{\widetilde{B}_x \to 0} \left( \langle S_z \rangle_{\mathrm{minima}_1} - \langle S_z \rangle_{\mathrm{minima}_2} \right) = 0 \tag{27}$$

The consequence of (27) for the experiment is that for small transversal $\widetilde{B}_x$-fields the non-harmonic dynamics is beyond the experimental resolution, but it could be useful to define a quasi-expectation value revival time (q-EVRT), which treats the MQT oscillation as an harmonic oscillation and defines the revival time $T_{rep}^{\mathrm{q\text{-}EVRT}}$ as the time between two successive minima of $\langle S_z \rangle_t$ (see figure C1). To estimate the quasi-EVRT in the tunneling regime with reasonable accuracy the standard ansatz of the time dependent perturbation theory can be used. Defining

$$|\Psi(t)\rangle = \sum_{n=0}^{2s} \sigma(t)_{s-n} e^{-iE_{(s-n)}t/\hbar} |\psi_{s-n}\rangle, \tag{28}$$

where $\sigma(t)$ are the time dependent coefficients and $E$ are the energy eigenvalues, and substituting (28), (26) in (1) we utilize the standard perturbation theory power series ansatz:

$$\sigma(t)_{s-n} = \sigma(t)_{s-n}^{(0)} + \lambda\sigma(t)_{s-n}^{(1)} + ... = \sum_{i=0}^{\infty} \lambda^i \sigma(t)_{s-n}^{(i)}. \tag{29}$$

Using coefficient comparison in first order perturbation theory, we obtain

$$\begin{aligned}
\sigma_s(t)^{(1)} &= \frac{\sigma_{s-1}(t_0)\widetilde{B}_x(S_x)_{1,2}}{E_{s-1}-E_s}\cdot(-e^{i(E_{s-1}-E_s)t/\hbar}+1)\\
&\vdots\\
\sigma_{s-n}(t)^{(1)} &= \widetilde{B}_x\Big(\frac{\sigma_{s-(n-1)}(t_0)(S_x)_{a,b}}{E_{s-(n-1)}-E_{s-n}}\cdot(-e^{i(E_{s-(n-1)}-E_{s-n})t/\hbar}+1)\\
&\quad+\frac{\sigma_{s-(n+1)}(t_0)(S_x)_{a,c}}{E_{s-(n+1)}-E_{s-n}}\cdot(-e^{i(E_{s-(n+1)}-E_{s-n})t/\hbar}+1)\Big),
\end{aligned} \tag{30}$$

where $(S_x)_{1,2}$ is an element of the $\hat{S}_x$ matrix from (12). The square of the absolute value of $\sigma_s(t)^{(1)}$ leads to the well known Fermi's golden rule which reads:

$$|\sigma(t)^{(1)}_s|^2 = \left|\sigma_{s-1}(t_0)\widetilde{B}_x(S_x)_{1,2}\right|^2\cdot\left|\frac{\sin\left(\frac{E_{s-1}-E_s}{2\hbar}t\right)}{(E_{s-1}-E_s)/2}\right|^2, \qquad |\sigma_{s-1}(t_0)|=1 \tag{31}$$

Equation (31) describes the probability of quantum transitions between neighboring states $|\psi_{\sigma(t)_s}\rangle$ and $|\psi_{\sigma(t)_{s-1}}\rangle$, and has a maximum for $E_s = E_{s-1}$, which is only valid for times $t$

$$|\sigma(t)^{(1)}_s|^2 \le 1 \quad\Longrightarrow\quad t < \frac{\hbar}{\left|\widetilde{B}_x(S_x)_{1,2}\right|}. \tag{32}$$

Comparing $t$ of (32) with $T_{MQT}$, obtained in numerical simulations, leads to a proportionality constant of $T_{MQT}/t \approx \pi/2$ within the tunneling regime, which can be expressed as

$$\lim_{\widetilde{B}_x\to 0}\frac{T_{MQT}}{t} = \frac{\pi}{2} \tag{33}$$

Furthermore, the analysis of the numerical data leads to

$$T_{MQT} \approx \frac{T^{\text{q-EVRT}}_{rep}}{2} \approx \frac{\pi\hbar}{2\left|\widetilde{B}_x(S_x)_{1,2}\right|}, \tag{34}$$

which occurs for a $\widetilde{B}_z/K = N \in \mathbb{Z}$ ratio with $N \le 2s$. Hence, the magnetization tunneling time $T_{MQR}$ at the avoided level crossing points is unambiguously related to the repetitive time $T^{\text{q-EVRT}}_{rep}$ if nonegligible transversal perturbations are present. In the limit of vanishing perturbations $T_{MQR} \to \infty$, that is, no tunneling occurs and (C.6) for $T^{\text{q-EVRT}}_{rep}$ can be applied. The conditions for magnetization tunneling $\widetilde{B}_z/K = 2s$ for the Hamiltonian of (26) is a subset of the condition $\widetilde{B}_z/K \in \mathbb{Z}$ defined in (21).

## 5. Conclusion

Analytical and numerical studies of the quantum dynamics of effective quantum spins have revealed that the quantum revival of expectation values and the total wave-function is identical for integer spin values, but very different for half-integer spins. This finding permitted to resolve the contradicting conclusions in the literature on the dependence of the revival time on the spin value. It has been concluded that the QRT doesn't depend on the spin number but only on the spin statistics (integer or half-integer). According to the derived analytical expressions the EVRT is shortest for integer field-anisotropy ratios. As the field can easily be tuned experimentally, and time-dependent measurements have become available in the last years, we hope that this finding will permit a highly presice measurement of magnetic anisotropies. For that purpose one should measure the EVRT as a function of external magnetic field and define the shortest one among all measured values.
An applied transverse field promotes the MQT. Our analysis shows that the EVRT is strongly correlated with the MQT. By increasing the revival time using specific combinations of material parameters and fields one can increase the life-times of the quantum states, which may be used as bits of information in future technologies.

## 6. Acknowledgement

We acknowledge financial support from the DPG in the framework of the SFB 668.

## Appendix A. Expectation values of spin $1\hbar$

The complete solution of the expectation values $\langle S_x \rangle, \langle S_y \rangle$ and $\langle S_z \rangle$ for an effective quantum spin $s = 1\hbar$ defined by the Hamilton operator $\hat{H} = -\hat{S}_z \widetilde{B}_z - K\hat{S}_z^2$ are:

$$
\begin{aligned}
\langle S_x \rangle &= \frac{2\hbar}{\sqrt{2}} \\
&\cdot \Big[ (\underset{real}{\varphi_{+1}}(t_0) \underset{real}{\varphi_0}(t_0) + \underset{imag}{\varphi_{+1}}(t_0) \underset{imag}{\varphi_0}(t_0)) \cdot \cos[(-B_z - \hbar K)t] \\
&+ (\underset{real}{\varphi_{-1}}(t_0) \underset{real}{\varphi_0}(t_0) + \underset{imag}{\varphi_{-1}}(t_0) \underset{imag}{\varphi_0}(t_0)) \cdot \cos[(B_z - \hbar K)t] \\
&+ (-\underset{real}{\varphi_{+1}}(t_0) \underset{imag}{\varphi_0}(t_0) + \underset{imag}{\varphi_{+1}}(t_0) \underset{real}{\varphi_0}(t_0)) \cdot \sin[(-B_z - \hbar K)t] \\
&+ (\underset{imag}{\varphi_{-1}}(t_0) \underset{real}{\varphi_0}(t_0) - \underset{real}{\varphi_{-1}}(t_0) \underset{imag}{\varphi_0}(t_0)) \cdot \sin[(B_z - \hbar K)t] \Big]
\end{aligned}
$$

$$
\begin{aligned}
\langle S_y \rangle &= \frac{2\hbar}{\sqrt{2}} \\
&\cdot \Big[ (\underset{real}{\varphi_{+1}}(t_0) \underset{real}{\varphi_0}(t_0) + \underset{imag}{\varphi_{+1}}(t_0) \underset{imag}{\varphi_0}(t_0)) \cdot \sin[(-B_z - \hbar K)t] \\
&+ (-\underset{real}{\varphi_{-1}}(t_0) \underset{real}{\varphi_0}(t_0) - \underset{imag}{\varphi_{-1}}(t_0) \underset{imag}{\varphi_0}(t_0)) \cdot \sin[(B_z - \hbar K)t] \\
&+ (-\underset{real}{\varphi_{+1}}(t_0) \underset{imag}{\varphi_0}(t_0) - \underset{imag}{\varphi_{+1}}(t_0) \underset{real}{\varphi_0}(t_0)) \cdot \cos[(-B_z - \hbar K)t] \\
&+ (\underset{imag}{\varphi_{-1}}(t_0) \underset{real}{\varphi_0}(t_0) - \underset{real}{\varphi_{-1}}(t_0) \underset{imag}{\varphi_0}(t_0)) \cdot \cos[(B_z - \hbar K)t] \Big]
\end{aligned}
\tag{A.1}
$$

$$
\langle S_z \rangle = \frac{\hbar}{2} \Big[ |\underset{real}{\varphi_{+1}}(t_0)|^2 - |\underset{real}{\varphi_{-1}}(t_0)|^2 + |\underset{imag}{\varphi_{+1}}(t_0)|^2 - |\underset{imag}{\varphi_{-1}}(t_0)|^2 \Big]
$$

These expressions have been obtained using $\langle \Psi(t) | \hat{S}_i | \Psi(t) \rangle$ $(i := x, y, z)$ where

$$
|\Psi(t)\rangle = \begin{pmatrix} \varphi_{+s}(t) = \varphi_{+1}(t_0) \cdot e^{-i(-\widetilde{B}_z - K)t/\hbar} \\ \varphi_0(t) = \varphi_0(t_0) \\ \varphi_{-1}(t) = \varphi_{-1}(t_0) \cdot e^{-i(\widetilde{B}_z - K)t/\hbar} \end{pmatrix}. \tag{A.2}
$$

## Appendix B. Fourier series Form

The expectation values $\langle S_x \rangle, \langle S_y \rangle$ and $\langle S_z \rangle$ for a general effective quantum spin $s$, defined by a Hamilton operator without off-diagonal elements, can be generalized as:

$$
\langle S_m \rangle = \langle \Psi(t) | S_m | \Psi(t) \rangle \qquad , m \in [x, y, z] \tag{B.1}
$$

where $|\Psi(t)\rangle$ is

$$\begin{pmatrix} \varphi_{+s}(t) = \varphi_{+s}(t_0) \cdot e^{-i(-s\widetilde{B}_z - s^2 K)t/\hbar} \\ \vdots \\ \varphi_{-s}(t) = \varphi_{-s}(t_0) \cdot e^{-i(s\widetilde{B}_z - s^2 K)t/\hbar} \end{pmatrix} \tag{B.2}$$

The generalized spin matrices $S_m$ are defined by

$$\begin{aligned} (S_{x_{a,b}}) &= \frac{\hbar}{2}\sqrt{(s+1)(a+b-1)-ab} \\ (S_{y_{a,b}}) &= \pm\frac{\hbar}{2i}\sqrt{(s+1)(a+b-1)-ab} \\ (S_{z_{a,a}}) &= \hbar(s-(a-1)) \\ \\ 1 \leq a&, b \leq 2s+1 \end{aligned} \tag{B.3}$$

The generalized components of the time dependent spin eigenstate vector $\left|\Psi_s(t)\right\rangle$ are

$$\begin{aligned} &\varphi_{s-j}(t) = \varphi_{s-j}(t_0) \cdot e^{-i(-(s-j)\widetilde{B}_z - (s-j)^2 K)\frac{t}{\hbar}} \\ \\ &0 \leq j \leq 2s \end{aligned} \tag{B.4}$$

The generalized form of $\left|\Psi_s(t)\right\rangle$ in (B.4) is a consequence of the fact that the Hamilton operator $\hat{H} = -\hat{S}_z\widetilde{B}_z - K\hat{S}_z^2$ does not contain off-diagonal elements. Hence, it represents a system of uncoupled linear ordinary differential equations of first order. Inserting (B.4) and (B.3) in (B.1) leads to

$$\begin{aligned} \left\langle S_m \right\rangle &= \varphi_s^*(t)(S_{m_{1,2}})\varphi_{s-1}(t) \\ &+ \varphi_{s-1}^*(t)\Big((S_{m_{2,1}})\varphi_s(t) + (S_{m_{2,3}})\varphi_{s-2}(t)\Big) \\ &+ \varphi_{s-2}^*(t)\Big((S_{m_{3,2}})\varphi_{s-1}(t) + (S_{m_{3,4}})\varphi_{s-3}(t)\Big) \\ &+ \varphi_{s-3}^*(t)\Big((S_{m_{4,3}})\varphi_{s-2}(t) + (S_{m_{4,5}})\varphi_{s-4}(t)\Big) \\ &\vdots \\ &+ \varphi_{-s}^*(t)(S_{m_{a,a-1}})\varphi_{-s+1}(t) \end{aligned} \tag{B.5}$$

We remark that the eigenstates and components $(S_{m_{a,a-1}})$ of the spin matrices $S_m$ in (B.5) have a systematic structure, which can be extracted in the following form:

$$\begin{aligned}\langle S_m \rangle &= \varphi_s^*(t)(S_{m_{1,2}})\varphi_{s-1}(t) + \varphi_{-s}^*(t)(S_{m_{a,a-1}})\varphi_{-s+1}(t)\\ &+ \sum_{j=1}^{2s-1} \Big(\varphi_{s-j}^*(t)\Big((S_{m_{j+1,j}})\varphi_{s-(j-1)}(t) + (S_{m_{j+1,j+2}})\varphi_{s-(j+1)}(t)\Big)\Big)\\ &, \quad m \in [x, y]\end{aligned} \tag{B.6}$$

Repeating and sorting of the (B.6) leads to

$$\langle S_z \rangle = \sum_{j=0}^{2s-1} \Big((S_{z_{j+1,j+1}})\varphi_{s-j}^*(t) \cdot \varphi_{s-j}(t)\Big) \tag{B.7}$$

and

$$\begin{aligned}\langle S_x \rangle_t &= \sum_{j=1}^{2s} \big(\alpha_j \cdot \cos(\omega_j t) + \beta_j \cdot \sin(\omega_j t)\big)\\ \alpha_j &= 2(S_x)_{j,j+1}\Big(\varphi(t_0)_{s-(j-1)}^{real} \cdot \varphi(t_0)_{s-j}^{real} + \varphi(t_0)_{s-(j-1)}^{imag} \cdot \varphi(t_0)_{s-j}^{imag}\Big)\\ \beta_j &= 2(S_x)_{j,j+1}\Big(-\varphi(t_0)_{s-(j-1)}^{real} \cdot \varphi(t_0)_{s-j}^{imag} + \varphi(t_0)_{s-(j-1)}^{imag} \cdot \varphi(t_0)_{s-j}^{real}\Big)\\ \langle S_y \rangle_t &= \sum_{j=1}^{2s} \big(\alpha_j \cdot \sin(\omega_j t) + \beta_j \cdot \cos(\omega_j t)\big)\\ \alpha_j &= 2(S_y)_{j,j+1}\Big(-\varphi(t_0)_{s-(j-1)}^{real} \cdot \varphi(t_0)_{s-j}^{real} - \varphi(t_0)_{s-(j-1)}^{imag} \cdot \varphi(t_0)_{s-j}^{imag}\Big)\\ \beta_j &= 2(S_y)_{j,j+1}\Big(\varphi(t_0)_{s-(j-1)}^{real} \cdot \varphi(t_0)_{s-j}^{imag} - \varphi(t_0)_{s-(j-1)}^{imag} \cdot \varphi(t_0)_{s-j}^{real}\Big)\\ \omega_j &= -\widetilde{B}_z - (2s - (2j-1))K\end{aligned} \tag{B.8}$$

Equation (B.8) is particularly interesting, because it has a Fourier series form. This form occurs if the Hamilton operator contains terms with an even exponent $\hat{S}^{2n}(n \in \mathbb{N} \setminus \{0\})$. It means the presented approach applies to all terms, which are quadratic in the spin operators, for example, different kinds of anisotropy.

## Appendix C. Derivation of critical $B_z/K$ ratios for EVRT

In this section we want to present a proof for the statement that the relation $\widetilde{B}_z/K = N \in \mathbb{Z}$ for constant $K$ always leads to a lower time $T_{rep}$ than that for $N \in \mathbb{Q} \setminus \mathbb{Z}$. We investigate the set of rational numbers $\mathbb{Q}$ only, because the "gcd(...)" used for the definition of quantum revival of expectation values (EVRT) given in the main text is only defined for integers. Due to the fact that the irrational numbers (subset of $\mathbb{R}$) can not be extended to become integers we can not use them for the "gcd(...)". First we formulate two Lemma's, which are necessary for our proof.

Lemma 1:
If $\gamma_1 + \gamma_2 = G$, in which $\gamma_1, \gamma_2, G \in \mathbb{Z}$, $\gamma_1/x \in \mathbb{Z}$, $G/x \in \mathbb{Z}$ and $x \in \mathbb{Q}$, then $\gamma_2/x \in \mathbb{Z}$. Otherwise it would lead to a contradiction. If only $G/x \in \mathbb{Z}$ is given, then $\gamma_1/x$ and $\gamma_2/x$ are not necessarily an element of $\mathbb{Z}$.

Proof 1. (Lemma 1):

$$\begin{aligned}
&G/x \in \mathbb{Z} \ \wedge \ \gamma_2/x \in \mathbb{Z} \\
&\frac{\gamma_1+\gamma_2}{x} = \frac{G}{x} \\
&\Rightarrow \frac{\gamma_1}{x} = \frac{G}{x} - \frac{\gamma_2}{x} \\
&\Rightarrow \frac{G}{x} - \frac{\gamma_2}{x} \in \mathbb{Z} \ \Rightarrow \frac{\gamma_1}{x} \in \mathbb{Z}
\end{aligned} \tag{C.1}$$

Proof 2. (Lemma 1):

$$\begin{aligned}
&G = 6 \ \wedge \ \gamma_1 = 2 \ \wedge \ \gamma_2 = 4 \wedge \ x = 3 \\
&\gamma_1 + \gamma_2 = G \ \Rightarrow 6 = 2 + 4 \\
&\Rightarrow \frac{G}{x} = 2 \ \in \mathbb{Z} \\
&\Rightarrow \frac{\gamma_1}{x} = \frac{2}{3} \notin \mathbb{Z} \\
&\Rightarrow \frac{\gamma_2}{x} = \frac{4}{3} \notin \mathbb{Z}
\end{aligned} \tag{C.2}$$

Lemma 2:
If $\omega_1 + \Delta_{\omega_1,\omega_2} = \omega_2$, in which $\omega_1, \omega_2, \Delta_{\omega_1,\omega_2} \in \mathbb{Z}$ then it follows for $\omega_2/x \in \mathbb{Z}$, $\omega_1/x \notin \mathbb{Z}$ and $\Delta_{\omega_1,\omega_2}/x \notin \mathbb{Z}$ that $\gcd(\omega_1, \omega_2) \neq x$.

Proof (Lemma 2):

$$\begin{aligned}
&\gcd(\omega_1, \omega_2) \leq \inf(\omega_1, \omega_2) \\
&\omega_1 < \omega_2 \\
&x \leq \omega_1 \\
&\omega_i/x \in \mathbb{Z} \ \ (\text{condition for } \gcd(\omega_1, ..., \omega_i)) \\
&\text{if } \frac{\omega_2}{x} \in \mathbb{Z} \ \wedge \ \frac{\omega_1}{x} \notin \mathbb{Z} \\
&\Rightarrow \gcd(\omega_1, \omega_2) \neq x
\end{aligned} \tag{C.3}$$

Lemma 1 and Lemma 2 enable us to use the structure of (C.8) to find all possible greatest common divisors.

$$\widetilde{B}_z = NK \tag{C.4}$$

$$\omega_j = -\widetilde{B}_z - (2s - (2j-1))K \tag{C.5}$$

$$\omega_j = \begin{cases} (-N-1)K, (-N-3)K, ..., (-N-(2n+1))K & \text{for integer } s \\ (-N)K, (-N-2)K, (-N-4)K, ..., (-N-2n)K & \text{for half-integer } s \end{cases} \tag{C.6}$$

$$n \in \mathbb{N}_0$$

Equation (C.6) demonstrates that the frequencies $\omega_j$ defined by (C.5) have the form of series.
Based on (C.6) we substitute $N$ via $N = a/b$, where $a, b \in \mathbb{Z}$. This substitution enables us to create any element of $\mathbb{Q}$. Further, we use the conclusion of the series form of (C.6) and exchange the numbers through a variable $\Gamma_n$, which contains all properties of (C.5) for $N = 0$.

$$N := \frac{a}{b}, \qquad N \in \mathbb{Q} \Rightarrow a, b \in \mathbb{Z} \tag{C.7}$$

$$\omega_j := \xi_n := \Big(\frac{a}{b} + \Gamma_n\Big)K, \qquad \Gamma_n = \begin{cases} 2n+1 & \text{for integer } s \\ 2n & \text{for half-integer } s \end{cases} \tag{C.8}$$

$$n \in \mathbb{N}_0 \tag{C.9}$$

Next, we expand the numerator and denominator of $T_{rep}$ with $b$ in order to obtain the necessary integer frequency condition for the greatest common divisor:

$$T_{rep} = \frac{2\pi}{\gcd(\xi_0, ..., \xi_j)} \quad \Rightarrow \quad T_{rep} = \frac{2\pi \cdot b}{\gcd(\xi_0 b, ..., \xi_j b)} \tag{C.10}$$

The expanded frequencies $\xi_n b$ in (C.10) are defined by

$$\xi_n b = \Big(a + \Gamma_n \cdot b\Big)K \tag{C.11}$$

Next, we substitute the first multiplier of (C.11) via $\Omega = 2m$ or $\Omega = 2m+1$ to distinguish between the even or odd solution for each frequency series $\xi_n b$:

$$\begin{aligned} \xi_0 b &= \Big(a + \Gamma_0 \cdot b\Big)K = \Omega K \\ &\vdots \\ \xi_n b &= \Big(a + \Gamma_n \cdot b\Big)K = (\Omega + n \cdot 2b)K \end{aligned} \qquad \Omega := \begin{cases} 2m \\ 2m+1 \end{cases} \tag{C.12}$$

$$\xi_n b - \xi_0 b = n \cdot 2b, \qquad m \in \mathbb{Z} \tag{C.13}$$

Equation (C.12) provides us with an expression which permits to use Lemma 1 and Lemma 2 in order to estimate the greatest common divisor.

As $\xi_0 b$ corresponds to the lowest frequency it follows that

$$\begin{aligned} &\gcd(\xi_0 b, ..., \xi_j b) \leq \xi_0 \cdot b \\ &\Rightarrow \gcd(\xi_0 b, ..., \xi_j b) \leq \Omega K \end{aligned} \tag{C.14}$$

In order to estimate if $\Omega K$ is a common divisor of all $\xi_n b$ we divide (C.12) by $\Omega K$, which leads to

$$\frac{\xi_n \cdot b}{\Omega K} = \Big(\frac{a}{\Omega} + \frac{\Gamma_n \cdot b}{\Omega}\Big) = \Big(1 + \frac{n \cdot 2b}{\Omega}\Big). \tag{C.15}$$

The $\Omega K$ in (C.15) is a common divisor of (C.12) for $\Omega \in [b, 2b, 1, 2]$. Next, we have to distinguish in which cases $\Omega K = bK$, $\Omega K = 2bK$, $\Omega K = 1K$ and $\Omega K = 2K$ is the greatest common divisor of (C.12). The case $\Omega = b$ leads to

$$\frac{\xi_n \cdot b}{\Omega K} = \Big(\frac{a}{b} + \Gamma_n\Big) \tag{C.16}$$

Because of $\Gamma_n \in \mathbb{Z}$ and Lemma 1 it follows from (C.16) that if $\xi_n \cdot b/\Omega K \in \mathbb{Z}$ than $a/b$ has to be an element of $\mathbb{Z}$. The case $\Omega = 2b$ leads to

$$\frac{\xi_n \cdot b}{\Omega K} = \Big(\frac{a}{2b} + \frac{\Gamma_n}{2}\Big), \tag{C.17}$$

and it follows that

$$\begin{aligned} &\Big(\frac{a}{2b} + \frac{\Gamma_n}{2}\Big) = \sigma \in \mathbb{Z} \\ &\Rightarrow \frac{a}{b} = 2\sigma - \Gamma_n \end{aligned} \tag{C.18}$$

Because of $\sigma,\ \Gamma_n \in \mathbb{Z}$ from (C.18) and Lemma 1 it follows that $a/b \in \mathbb{Z}$. The conclusion from (C.16) and (C.17) is that

$$\gcd(\xi_1 \cdot b, ..., \xi_n \cdot b) = \begin{cases} 2bK \\ bK \end{cases} \tag{C.19}$$

for $a/b \in \mathbb{Z}$, which leads to

$$T_{rep} = \begin{cases} \dfrac{2\pi}{K} \\ \dfrac{\pi}{K} \end{cases} \tag{C.20}$$

Because of the result from (C.20) and Lemma 2 it follows by insterting $\Omega = 1$ and $\Omega = 2$ in (C.15) that the greatest common divisor of (C.12) has to be $\Omega K = 1K$ or $\Omega K = 2K$ for $a/b \notin \mathbb{Z}$, which leads to:

$$T_{rep} = \begin{cases} \dfrac{\pi b}{K} & ,\ b > 2 \\ \dfrac{2\pi b}{K} & ,\ b > 1 \end{cases} \tag{C.21}$$

The estimation of the lower bound for the $b$ in (C.21) is done by (C.12) by permutation of even and odd $\Omega$.
The consequence of (C.21) and (C.20) is that

$$T_{rep}^{N\in\mathbb{Z}} < T_{rep}^{N\notin\mathbb{Z}}. \tag{C.22}$$

q.e.d.

Figure C1 represents a visualization of the result of (C.22).

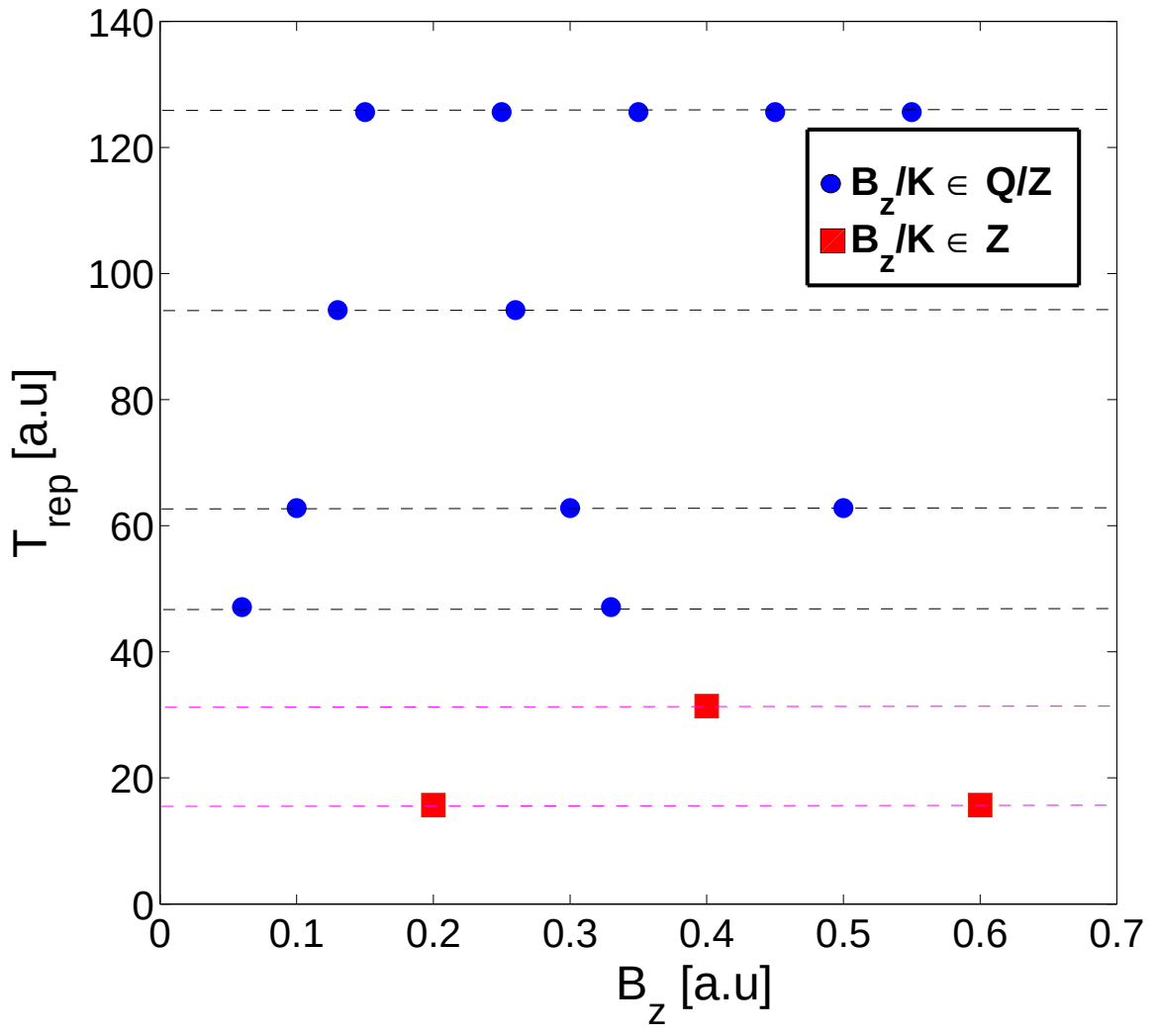

**Figure C1.** Revival times of expectation values (EVRT) of a spin $1\hbar$ for different $B_z$ values and $K = 0.2$ [a.u] = const. Because the EVRT is not continuous for any $B_z$ we choose to give an example for discrete $B_z$ values. This demonstrates the result $T_{rep}^{N\in\mathbb{Z}} < T_{rep}^{N\notin\mathbb{Z}}$.

## Appendix D. Derivation of critical field-anisotropy ratios for QRT

In this section we derive the QRT for two scenarios: $\widetilde{B}_z/K = N \in \mathbb{Z}$ and $\widetilde{B}_z/K = N \notin \mathbb{Z}$ with $K = const$. We start with the wave functions from the chapter B

$$\begin{aligned} \varphi_{+s}(t) &= \varphi_{+s}(t_0) \cdot e^{-i(-s\widetilde{B}_z - s^2K)\frac{t}{\hbar}} \\ &\vdots \\ \varphi_{-s}(t) &= \varphi_{-s}(t_0) \cdot e^{-i(s\widetilde{B}_z - s^2K)\frac{t}{\hbar}} \end{aligned} \tag{D.1}$$

where

$$\omega_j = \frac{(-s-j)\widetilde{B}_z - (s-j)^2K}{\hbar}, \qquad \text{for} \quad j \in \mathbb{N}_0 \tag{D.2}$$

$$\omega_j = \begin{cases} (-N-1)K, (-2N-4)K, (-3N-9)K, ..., (-nN-n^2)K & \text{for integer } s \\ (-2N-1)K, (-6N-9)K, ..., (-(4n+2)N-(2n+1)^2)K & \text{for half-integer } s \end{cases} \tag{D.3}$$

$$n \in \mathbb{N}_0$$

The derivation follows the same principles like in Appendix C. Lemma 1 and Lemma 2 are guidelines for the following derivation.

Integer $s$ case:

$$\begin{aligned} \xi_0 b &= (a+b)K = \Omega K \\ \xi_1 b &= \Big(2(a+b)+2b\Big)K = (2\Omega+2b)K \\ &\vdots \\ \xi_n b &= \Big((n+1)(a+b)+b\sum_{i=0}^{n} 2n\Big)K = \Big((n+1)\Omega + b\sum_{i=0}^{n} 2n\Big)K \end{aligned} \qquad \Omega := \begin{cases} 2m \\ 2m+1 \end{cases} \tag{D.4}$$

$$\frac{\xi_1 b}{\Omega} = (2+\frac{2b}{\Omega})K \tag{D.5}$$

For $a/b = N \in \mathbb{Z}$ it follows that:

$$T_{rep} = \begin{cases} \frac{2\pi}{K} \\ \frac{\pi}{K} \end{cases} \tag{D.6}$$

For $a/b = N \notin \mathbb{Z}$ (D.4) leads for the greatest common divisor to

$$\mathrm{gcd}(\xi_0 b, ..., \xi_j b) = \begin{cases} 2K \\ K \end{cases} \tag{D.7}$$

$$\Rightarrow T_{rep} = \begin{cases} \frac{\pi b}{K} & , b > 2 \\ \frac{2\pi b}{K} & , b > 1 \end{cases} \tag{D.8}$$

Half-Integer $s$ case:

$$\xi_0 b = (2a+b)K = \Omega K$$
$$\xi_1 b = \Big(3(2a+b)+6b\Big)K = (3\Omega+6b)K \qquad \Omega := \begin{cases} 2m \\ 2m+1 \end{cases}$$
$$\vdots$$
$$\xi_n b = \Big((2n+1)(2a+b) - 2nb + b\sum_{i=0}^{n} 8n\Big)K = \Big((2n+1)\Omega - 2nb + b\sum_{i=0}^{n} 8n\Big)K \tag{D.9}$$

For $N \in \mathbb{Z}$ it follows

$$\frac{\xi_1 b}{\Omega} = \Big(3 + \frac{6b}{\Omega}\Big)K$$
$$\frac{\xi_2 b}{\Omega} = \Big(5 + \frac{20b}{\Omega}\Big)K \tag{D.10}$$

The result of (D.10) is that

$$\gcd(\xi_0 b, ..., \xi_j b) = bK$$
$$\Rightarrow T_{rep} = 4\cdot\frac{2\pi}{K} \qquad \text{for} \quad N \in \mathbb{Z} \tag{D.11}$$

For $N \notin \mathbb{Z}$ one can write

$$\frac{\xi_1 b}{\Omega} = \Big(3 + \frac{6b}{\Omega}\Big)K = \Big(3 + \frac{6b}{2a+b}\Big)K$$
$$\frac{\xi_2 b}{\Omega} = \Big(5 + \frac{20b}{\Omega}\Big)K = \Big(5 + \frac{20b}{2a+b}\Big)K \tag{D.12}$$

$$\frac{6b}{2a+b} = \beta \in \mathbb{Z} \quad \Rightarrow \frac{a}{b} = \frac{(6-\beta)}{2\beta}$$
$$\frac{20b}{2a+b} = \beta \in \mathbb{Z} \quad \Rightarrow \frac{a}{b} = \frac{(20-\beta)}{2\beta} \tag{D.13}$$

$$\frac{(6-\beta)}{2\beta} = \frac{(20-\beta)}{2\beta}$$
$$\Rightarrow 14 = 0 \quad \Rightarrow \quad \gcd(\xi_0 b, ..., \xi_j b) \neq \Omega K \tag{D.14}$$

The next step is to estimate which of the values $2bK$ or $bK$ is the greatest common divisor. Because of the condition $(2a+b)/2b \in \mathbb{Z}$ for $a = \frac{b(2\gamma-1)}{2}$ it follows that

$$\begin{aligned} &\gcd(\xi_0 b, ..., \xi_j b) = 2K \\ &\Rightarrow T_{rep} = 4 \cdot \frac{\pi}{K} \qquad \text{for} \;\; N = \frac{(2\gamma - 1)}{2} \notin \mathbb{Z} \end{aligned} \tag{D.15}$$

For $N \neq (2a + b)/2b$ and $N \notin \mathbb{Z}$ it follows

$$\Rightarrow \gcd(\xi_0 b, ..., \xi_j b) = \begin{cases} 2K \\ K \end{cases} \tag{D.16}$$

$$\Rightarrow T_{rep} = \begin{cases} 4 \cdot \dfrac{2\pi b}{K} & , \; b > 2 \\ 4 \cdot \dfrac{\pi b}{K} & , \; b > 3 \end{cases} \tag{D.17}$$

We can summerize the results for the half-integer $s$ case as:

$$T_{rep} = \begin{cases} 4 \cdot \dfrac{2\pi}{K} & , \; N \in \mathbb{Z} \\ 4 \cdot \dfrac{\pi}{K} & , \; N = \dfrac{(2\gamma - 1)}{2} \notin \mathbb{Z} \\ 4 \cdot \dfrac{2\pi b}{K} & , \; b > 2 \quad N \notin \mathbb{Z} \\ 4 \cdot \dfrac{\pi b}{K} & , \; b > 3 \quad N \notin \mathbb{Z} \end{cases} \tag{D.18}$$

By comparing (D.18) with (C.21) and (C.20) from Appendix C for the same values of $K$, $a$, and $b$ it follows:

$$T_{rep}^{\mathrm{EVRT}} = \frac{1}{\alpha} \cdot T_{rep}^{\mathrm{QRT}}, \quad \alpha \in [1, 2, 4, 8], \qquad \text{for half-integer } s \tag{D.19}$$

We use this expression in the main text.

[1] Eberly J H, Narozhny N B and Sanchez-Mondragon J J 1980 Phys. Rev. Lett. **44** 1323

[2] Seltzer S J, Meares P J and Romalis M V 2007 Phys. Rev. A **75** 051407R

[3] Robinett R W 2004 Phys. Rep. **392** 1

[4] Schmidt A G M, Azeredo A D and Gusso A 2008 Phys. Lett. A **372** 16

[5] Gora P F and Jedrzejek C 1993 Phys. Rev. A **48** 3291

[6] Narozhny N B, Sanchez-Mondragon J J and Eberly J H 1981 Phys. Rev. A **23** 236

[7] Fleischhauer M and Schleich W P 1993 Phys. Rev. A **47** 4258

[8] Bluhm R, Kostelecky A and Porter J 1996 Am. J. Phys. **64** 944

[9] Agarwal G S and Banerji J 1998 Phys. Rev. A **57** 3880

[10] Aronstein D L and Stroud C R 1997 Phys. Rev. A **55** 4526

[11] Aronstein D L and Stroud C R 2000 Phys. Rev. A **62** 022102

[12] Styer D F 2001 Am. J. Phys. **69** 56

[13] Wright E M, Walls D F and Garrison J C 1996 Phys. Rev. Lett. **77** 2158

[14] Villain P and Lewenstein M 2000 Phys. Rev. A **62** 043601

[15] Gauyacq J P and Lorente N 2013 Phys. Rev. B **87** 195402

[16] Thiele S, Balestro F, Ballou R, Klyatskaya S and Ruben M 2014 Science **344** 1135

[17] Gatteschi D and Sessoli R 2003 Angew. Chem. Int. Ed. **42** 3

[18] Khajetoorians A A, Wiebe J, Chilian B, Lounis S, Blügel S and Wiesendanger R 2012 Nature Phys. **8** 497

[19] Zhou L, Wiebe J, Lounis S, Vedmedenko E Y, Meier F, Blügel S, Dederichs P H and Wiesendanger R 2010 Nature Phys. **6** 187

[20] Wieser R 2011 Phys. Rev. B **84** 054411

[21] Gauyacq J P and Lorente N 2014 Surf. Sci. **630** 325

[22] Them K, Stapelfeldt T, Vedmedenko E Y and Wiesendanger R 2013 New J. Phys. **15** 013009

[23] Neumann A, Altwein D, Thönnißen C, Wieser R, Berger A, Meyer A, Vedmedenko E Y and Oepen H P 2014 New J. Phys. **16** 083012

[24] Piquerel R, Gaier O, Bonet E, Thirion C and Wernsdorfer W 2014 Phys. Rev. Lett. **112** 117203

[25] Bocchieri P and Loinger A 1957 Phys. Rev. **107** 2

[26] Loth S, Etzkorn M, Lutz C P, Eigler D M and Heinrich A J 2010 Science **329** 1628

[27] Wernsdorfer W 2008 Comptes Rendus Chimie **11** 10

[28] Cheng-Wei Huang W and Batelaan H 2013 J. Comp. Meth. Phys. **2013** 308538

[29] Bullynck M 2009 Historia Mathematica **36** 1

[30] Gatland I R 1991 Am. J. Phys. **59** 155

[31] Zener C 1932 Proc. R. Soc. London Ser. A **137** 696

[32] Thorwart M, Grifoni M and Hänggi P 2000 Phys. Rev. Lett. **85** 860